\documentstyle[12pt,epsfig]{article}

\graphicspath{fig/}

\begin{document}
\title
{\Large  FIELD OF THE CURRENT PULSE MOVING \\ ALONG
 THE STRAIGHT LINE WITH SUPER LIGHT VELOCITY.}
\author
{{\bf F.F.\,Valiev}\\
 V.A. Fock Institute for Physics,
St.Petersburg State Univ., \\ St.Petersburg, 198504, Russia \\}
\maketitle \vspace{0.3cm} \noindent{\bf {\it
Abstract\,\,\,}}{\small{{\bf A simulation of electric current
pulses formed by a packet of gamma-quanta moving through an
absorptive medium is presented. The electromagnetic fields of the
current pulse moving along the straight line with super light
velocity are obtained.\vspace{3ex}}}}

\section{Introduction}

In recent years the electromagnetic fields of sources moving with
velocities equal or higher than velocity of light \textbf{c} in
vacuum have been discussed in connection with an interest to the
problem of localized electromagnetic waves \cite{Recami,Borisov}.
Such a source can be obtained by interaction of a directed beam of
a hard electromagnetic radiation (X-rays) with an absorptive rod.
It is important to have the length of this rod larger than its
diameter. If initial quanta are moving along the axis of this rod,
the electron's current pulse whose velocity is equal to \textbf{c}
is generated. If the beam is directed at an angle to the rod's
axis, the current pulse has a super light velocity. The velocities
of electrons, of course, are smaller than \textbf{c}, but the
motion of area, where the current's density is non zero, can be
presented as the motion of some effective charge which is faster
than the velocity of light in vacuum. The investigation of the
spatial structure and of the time dependence of secondary
electromagnetic fields, generated by such a pulse, requires
solution of two problems. Firstly, we have to find the shape of
the current pulse generated by the beam of primary hard
electromagnetic radiation. Secondly, we have to solve the
Maxwell's equation for this current's pulse with zero initial
conditions. The current pulse in question was earlier studied
mainly on a phenomenological basis. The current pulse moving with
velocity of light was considered in connection with nuclear
explosions \cite{Karzas}.
\newline In this paper we
calculate the ``realistic'' shape of current pulse moving with
super-light velocity, taking into account the microscopic
movements of the electrons. We obtained electromagnetic fields
with this current pulse by using the approach presented in [4],
based on the method of an incomplete separation of variables by
V.I. Smirnov \cite{Smirnov} and Riemann formula. \vspace{3ex}

\section{The shape of the super-luminal current \\ pulse generated by a
shot-lived beam of a hard radiation}

The solution of the first problem of finding the shape of current
pulse requires to take into account the photon absorption's
processes, Compton scattering, and secondary coupling effects of
``delta electrons'' with matter which result in the formation of
the ions and the electrons. We found the shape of current pulse
via the computational experiment using the method of numerical
modelling described in \cite{Valiev}. This method enables us to
study the time dependence of the electrons' distribution in the
phase space. The code GEANT \cite{GEANT} was used. The generation
of the electrons with energy lower than 10 keV was not taken into
account.

The scheme and the initial conditions of the computational
experiment are as follows. The absorbing area is placed in vacuum
and is bounded by the cylindrical surface of 0.3 m. in diameter
and of 50 m. in length L and by the two planes, which are
orthogonal to the cylinder axis. Its walls are made of
hydrocarbon-based material. Their thickness is 0.001 m. The
absorptive medium is air at the pressure of 1 atm. The symmetry
axis of the absorbing area coincides with the axis \textbf{z} of a
cylindrical coordinate system .

The origin of the coordinate system is chosen to be localized at
the point O, which lies in one of the two above mentioned planes.

The packet of the primary photons (a gamma-ray pulse) starts from a surface
formed by rotation of a line segment, which is at an angle $\pi $/2- $\theta
$ to the Z axis . The photons are simultaneously emitted at an angle $\theta
$ to the direction of the z axis.

The computational experiment is beginning with the starting time of the
gamma-ray pulse. The results of computer experiments are as follows:

1. We obtained the distribution \textbf{N($\rho )$} of the secondary
photoelectrons which have passed through the plane fixed at \textbf{z=z}$%
_{0} $. This distribution describes the electron localization in the radial
direction \textbf{$\rho $}.

2. The width ($\Delta $R) of the distribution N\textbf{($\rho )$} calculated
for the above initial conditions is less than 30 centimeters ($\Delta R< 30
cm$). The obtained value meets the necessary requirement of the
applicability of the model of a linear current ($L >>\Delta R$), used in the
simplified electrodynamic calculations.

The time dependence of the current's pulse passing through the
absorptive area at $z_{0}=30$ meters is shown in Fig. 1.

\bigskip
\begin{figure}[h]
\centerline{\epsfig{file=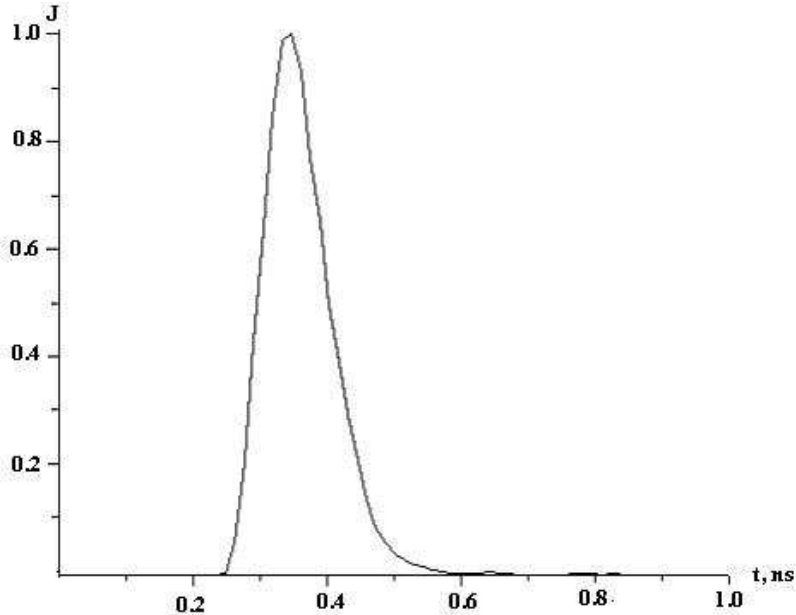,width=12cm}} \caption[dummy]
{The normalized current pulse in the cross section of the
absorbing region at z = 30 m.}
\end{figure}
\bigskip

In this plot the value of function
in maximum is presented as 1.0 on the axis of ordinates. The pulse length ($%
\Delta $T) taken on the ordinates level of 0.1 does not exceed 0.2 ns (or 6
cm in \textbf{ct} - units). The obtained value of $\Delta $T estimates the
localization of the current along \textbf{z }-- axis.

The shape of the running pulse can be described by the following function:
\begin{equation}
J(z,\tau )=A\cdot (1-\exp (-\alpha _{1}(\beta \tau -z)))\cdot \exp
(-\alpha _{2}(\beta \tau -z))
\end{equation}
$\alpha _{i}>0 $, i=1,2,

\noindent where $\tau =ct$ is the time variable, $\beta =v/c$ ,
$A$ is constant ; coefficients $\alpha _{i}$ are obtained by the
least square method

Thus, as a result of the above computational experiment the
space-time description of the hard radiation inducing super light
running pulse was obtained. \vspace{3ex}

\section{ Calculation of the magnetic component of the
electromagnetic wave generated by the current pulse moving with
super light velocity}
The basic solution of the Maxwell equations
that describes waves generated by a current pulse propagating
along the section of a strait line was given in [2]. In that work
the components of both the electric and magnetic fields are
expressed via a single scalar function $v(\rho,z , \tau)$. In the
present work we represent only magnetic component B. That was made
to simplify the subsequent physical interpretation of the obtained
results.

In our case the parameters of the hard radiation and absorptive medium,
geometry of the absorptive area were chosen in accordance with the condition
$L>>\Delta R$. The boundary effects in the description of a source of
electrodynamics problem are not taken into account (the necessary
requirement $L>> \Delta T$ is met). This means that for the calculation of
the required electromagnetic field (its magnetic component) we may use the
simplified model of a linear current.

Thus the source in the electrodynamics problem (non-zero component of the
current's density vector) can be taken in the form:

\begin{equation}
j_{z}={\frac{{\delta (\rho )}}{{2\pi \rho }}}h(\beta \tau
-z)h(z)h(L-z)J(z,\tau )\quad \tau >0\quad _{,}
\end{equation}
$$
 j_{z}=0\quad
\tau <0 $$ \ Here $\delta (\rho )$ is the Dirac distribution, h(z)
is the Heaviside step function. \ In the case of the axial -
symmetric source of an electrodynamics problem, the components of
the intensity \textbf{E} of electric field and the induction
\textbf{B} of magnetic field may be expressed via one scalar
function $v$ according to the following expressions:

\begin{equation}
E_{\rho }={\frac{{\partial ^{2}u}}{{\partial \rho \partial z}}}\quad E_{z}=-{%
\frac{{\partial ^{2}u}}{{\partial \tau ^{2}}}}+{\frac{{\partial ^{2}u}}{{%
\partial z^{2}}}}\quad B_{\varphi }={\frac{{\partial ^{2}u}}{{\partial \tau
\partial \rho }}}\quad _{,}  \label{eq1}
\end{equation}
\ Now the electrodynamics problem can be expressed in the form:

\begin{equation}
\left( {\frac{\partial ^{2}}{\partial \tau ^{2}}-\frac{1}{\rho }\frac{%
\partial }{\partial \rho }\left( {\rho \frac{\partial }{\partial \rho }}%
\right) -\frac{\partial ^{2}}{\partial z^{2}}}\right) v=\frac{4\pi }{c}%
j_{z},j_{z}=0\tau <0_{,}
\end{equation}

\bigskip \noindent were $v=\partial u / \partial \tau$ \ The
non-zero component of the magnetic induction $B_{\varphi }$vector is equal to%
$-\partial v/\partial \rho $.

The solution of the problem (4) in space - time representation is defined by
the following expression:

\begin{eqnarray}
&& v(\rho ,z,\tau )={\frac{1}{c}}\int\limits_{0}^{\tau }{d\tau }
\int\limits_{\tau ^{\prime }+z-\tau }^{-\tau ^{\prime }+z+\tau
}{dz^{\prime }h(z^{\prime })}  \\
&& h(\beta \tau ^{\prime }-z^{\prime })h(L-z^{\prime })J(z^{\prime
},\tau ^{\prime }){\frac{{\delta (\tau ^{\prime }-\tau +\sqrt{
\rho ^{2}+(z-z^{\prime })^{2}})}}{\sqrt{\rho ^{2}+(z-z^{\prime
})^{2}}}} \nonumber
\end{eqnarray}

After interchanged the order of an integration, we receive:

\begin{eqnarray}
v(\rho ,z,\tau ) ={\frac{1}{c}}\int\limits_{z-\tau }^{z}{dz^{\prime }}%
h(z^{\prime })h(L-z^{\prime }){\frac{1}{\sqrt{\rho ^{2}+(z-z^{\prime })^{2}}}%
}I_{1}  \nonumber \\
\quad +{\frac{1}{c}}\int\limits_{z}^{z+\tau }{dz^{\prime
}}h(z^{\prime })h(L-z^{\prime }){\frac{1}{\sqrt{\rho
^{2}+(z-z^{\prime })^{2}}}}I_{2}
\end{eqnarray}
\newline
where

\[
I_1 = \int\limits_0^{ - z^{\prime}+ \tau - z} {d\tau
^{\prime}J(z^{\prime},\tau ^{\prime})h(\beta \tau ^{\prime}-
z^{\prime})\delta (\tau ^{\prime}- \tau + \sqrt {\rho ^2 + (z -
z^{\prime})^2} )}
\]

\[
I_2 = \int\limits_0^{ - z^{\prime}+ \tau + z} {d\tau
^{\prime}J(z^{\prime},\tau ^{\prime})h(\beta \tau ^{\prime}-
z^{\prime})\delta (\tau ^{\prime}- \tau + \sqrt {\rho ^2 + (z -
z^{\prime})^2} )}
\]
\newline
For a super light delta - pulse
$$j_{z}={\frac{{\delta (\rho )}}{{2\pi \rho }}%
}\delta (\beta \tau -z)h(z)h(l-z)f(z,\tau )$$
 \newline (where $f$
is differentiable function) expression (6) can be simplified

\[
v(\rho ,z,\tau )={\frac{1}{c}}(\psi _{1}+\psi _{2})
\]
\newline

Here
\begin{eqnarray}
&&\psi _{1}(\rho ,z,\tau ) =                         \nonumber \\
&& f \left(\beta \left(\frac{\beta z-\tau+\sqrt{D}}{\beta
^{2}-1}\right), \frac{\beta z-\tau +\sqrt{D}} {\beta^{2}-1}\right)
\frac{1-h(\sqrt{D}+(1-\beta ^{2})\tau
 -\beta (z-\beta \tau ))}{\sqrt{D}}                    \nonumber \\
&&\times \{[h(L)(h(z-\tau )-h(z+\tau))]                  \nonumber \\
&&\times [h(\beta (\sqrt{D}-z\beta +\tau ))
-h(-z\beta^{2}+L(-1+\beta^{2})+\beta (\sqrt{D}+\tau ))] \nonumber\\
&&+
h(L-z-\tau )h(z+\tau ) \\
&&\times [h(-z+\sqrt{D}\beta +(-1+\beta +\beta ^2)\tau
)-h(-z{\beta^{2}+L(-1+\beta ^{2})+\beta (\sqrt{D}+\tau )})] \nonumber\\
&&+
h(z-\tau )h(L-z+\tau ) \nonumber\\
&&\times [-h(-z+\sqrt{D}\beta +(1+\beta -{\beta ^{2}})\tau
)+h(-z\beta ^{2}+L(-1+\beta ^{2})+\beta (\sqrt{D}+\tau
))]\}\nonumber
\end{eqnarray}
,

\begin{eqnarray}
&&\psi _{2}(\rho ,z,\tau )= \nonumber\\
&&\times f \left(\beta \left(\frac{\beta z-\tau -\sqrt{D}}{\beta
^{2}-1}\right),\frac{\beta z-\tau -\sqrt{D}}{\beta ^{2}-1}\right)
\frac{1-h(\sqrt{D}+(1-\beta ^{2})\tau +
\beta (z-\beta {\tau }))}{\sqrt{D}} \nonumber \\
&&\times  \{[h(L)(h(z-\tau )-h(z+\tau ))]\nonumber \\
&&\times  [ h(-\beta (\sqrt{D}+z\beta -\tau ))-h(-z\beta
^{2}+L(-1+\beta
^{2})+(\beta -\sqrt{D}+\tau ))] \\
&&+
h(L-z-\tau )h(z+\tau )\nonumber \\
&& \times  [h(-z+\sqrt{D}\beta -(-1+\beta +\beta
^{2}))\tau-h(-z\beta
^{2}+L(-1+\beta ^{2})+\beta (-\sqrt{D}+\tau ))]\nonumber \\
&&+
 h(z-\tau )h(L-z+\tau ) \nonumber \\
&&\times [-h(-z+\sqrt{D}\beta +(-1-\beta +\beta ^{2})\tau
)+h(-z{\beta ^{2}+L(-1+\beta ^{2})+\beta (-\sqrt{D}+\tau )})]\}
\nonumber,
\end{eqnarray}

where $D = (z - \beta \tau )^2 - \rho ^2(\beta ^2 - 1)$. This
expression may be used also for the nondispersive matter ($\beta $
is the ratio of velocity of the source motion to the light
velocity in given environment in this case).

We carried out the numerical calculation of super light source
using the expression (6). The time dependence of $v(\rho ,z,\tau
)$ is presented on fig. 2 for the seven angles of gamma - quanta.

\bigskip
\begin{figure}[h]
\centerline{\epsfig{file=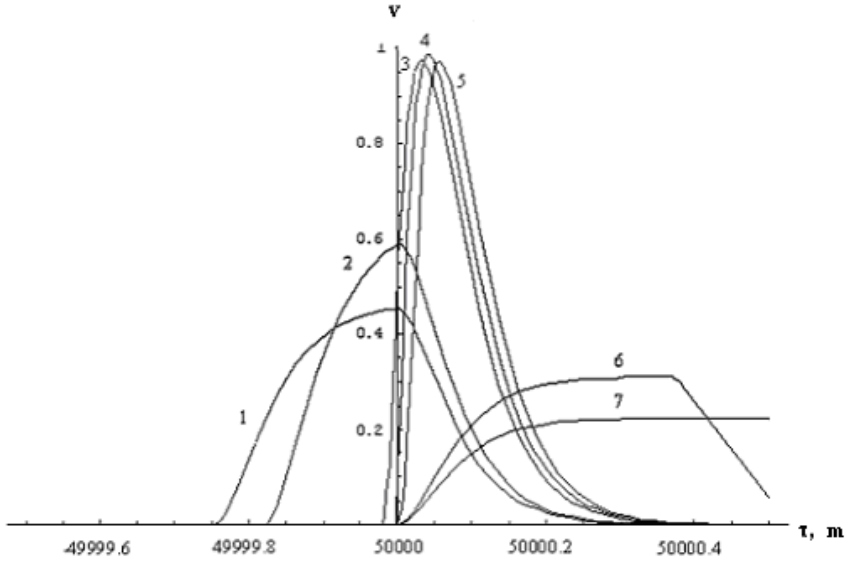,width=12cm}} \caption[dummy]
{Time dependence (in $ct$ units) of $v(\rho ,z,\tau )$ normalized
to
the maximal amplitude of different observation angles $\theta =0.3^{0}$(1), $%
\theta =3^{0}$(2), $\theta =5.5^{0}$,$\theta =5.44^{0}$(1),
$\theta =\arccos
(c/v)=arctg(0.1)\approx 5.71^{0}$(4), $\theta =6^{0}$(5),$\theta =9^{0}$(6),$%
\quad \theta =10^{0}$(7)$_{{}}$}
\end{figure}
\bigskip

In the proximity of the angle
$\theta =\arccos (c/v)=5.71^{0}$ the function $v(\rho ,z,\tau )_{{}}$%
sharpens its form. Note, that the variation of parameters of
numerical experiment causes essential changes of $v(\rho ,z,\tau
)$ . In particular, the reduction of value of $L$ results in
disappearance of a characteristic maximum in the distribution
$v(\theta )$ at $\theta =\arccos (c/v)$. The dependence of
$B_{\varphi }$ component on $\tau $ at various angles of
observation $\theta $ ($L=50 m$, $Z=50000 m$) is presented in
Fig.3.

\bigskip
\begin{figure}[h]
\centerline{\epsfig{file=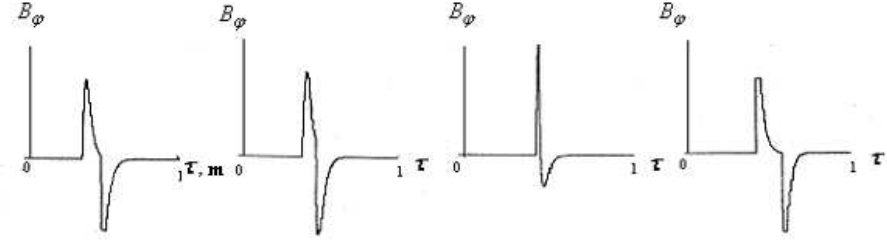,width=12cm}} \caption[dummy]
{$B_{\varphi }$ components of magnetic induction at various angles
of observation $\theta =0.3^{0}$(1), $\theta =3^{0}$(2) , $\theta
=\arccos
(c/v)=arctg(0.1)\approx 5.71^{0}$, $\theta =6^{0}$(4),$\theta =9^{0}$(5)$%
\quad \theta =10^{0}$(6) and $L=50 m$, $Z=50000 m$}
\end{figure}
\bigskip

As we see, duration of the two-polar pulses of an electromagnetic
field depends on angle $\theta $. The reduction of duration and
the growth of amplitude of the first half-wave in the vicinity of
$\theta =\arccos (c/v)$ is observed. The representation
$B_{\varphi }(\rho ,z,\varphi ,\tau )$ by Fourier integral allowed
us to obtain the frequency distribution of energy of the
electromagnetic field for different values of observation angles
in a far zone. The frequency spectra of electromagnetic pulses for
observation angles
: $\theta =0.3^{0}$, $\theta =3^{0}$ , $\theta =5.5^{0}$ ,$\theta =5.44^{0}$%
, $\theta =\arccos (c/v)=arctg(0.1)\approx 5.71^{0}$, $\theta =6^{0}$,$%
\theta =9^{0}\quad \theta =10^{0}$ in the optical range are
presented in Fig. 4.

\bigskip
\begin{figure}[h]
\centerline{\epsfig{file=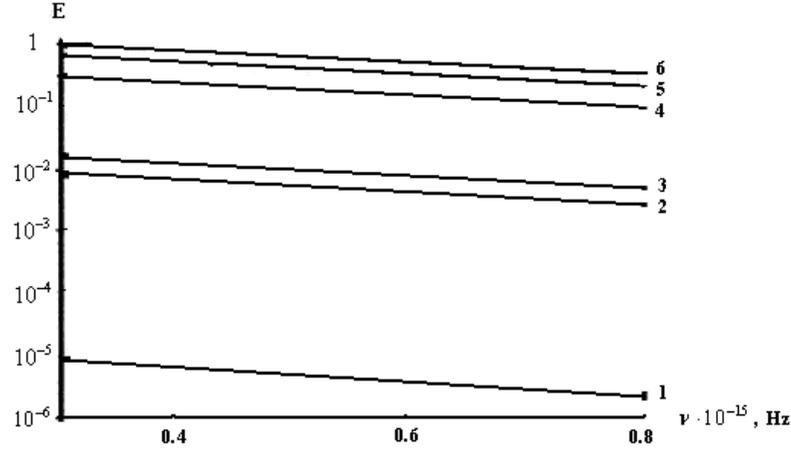,width=12cm}} \caption[dummy]
{The energy spectra of electromagnetic pulse observed at various
angles $_{{}}\theta =6.4^{0}$(1), $\theta =0.3^{0}$(2), $\theta =3^{0}$ , $%
\theta =5.5^{0}$ (4) , $\theta =5.8^{0}$(5), $\theta =\arccos
(c/v)=arctg(0.1)\approx 5.71^{0}$(6) in the far zone ( $Z=50000m)$
in optical frequency range ($3\ast 10^{14}-8\ast 10^{14})$ Hz for
$v/c\cong 1.005$. }
\end{figure}
\bigskip

As we see, radiation at the angle $\theta =\arccos (c/v)$ to the
direction of current pulse motion in the visible range of a
spectrum has maximal value. \vspace{3ex}

\section{ Summary}

We have calculated the shape of a the current pulse moving with velocity
greater then the velocity of light in vacuum for the certain geometry of the
hard radiation at fixed values of parameters of gamma - quanta pulse moving
through a layer of air.

The analytic representation for the potential $v(\rho ,z,\tau )$ was
obtained for the case of the superluminal delta pulse.

The spectra of radiation in the optical range for some different
observation angles range for $v/c$ equal 1.005 were calculated. We
found that the radiation has maximum at $\theta \cong \arccos
(c/v)$. 


\vspace{3ex}

\begin{thebibliography}{99}

\bibitem{Recami}{\small{\small E. Recami .  Foundations of Physica  \textbf{31}.No.7
1119(2001).

\bibitem{Borisov} Borisov V.V., Utkin A.B.    J.Phys.D.: Appl.Phys
 614(1995)

\bibitem{Karzas}  Karzas W.J., Letter R
 Phys.Rev. B.,  , \textbf{137}(5) ,
1369(1965).

\bibitem{Manankova} Manankova A.V.
Izv.Vyssh.Uchebn.Zaved.Radiofiz,\textbf{15}, 211(1972).


\bibitem{Smirnov} V.I. Smirnov  {\it A Course of Higher Mathematics, Vol.4.} (Nauka
press,Moskow,1965).

\bibitem{Valiev} F.F.Valiev
Tech. Phys.,\textbf{46}. No.12  1579(2001).


\bibitem{GEANT}  User's Guide. CERN DD/EE/83-1 (1983)}}

\end{thebibliography}
\end{document}